\documentclass[letter]{aa}

\usepackage[varg]{txfonts}
\usepackage{graphicx}
\usepackage[colorlinks=true,
    linkcolor=blue, citecolor=blue, filecolor=blue, urlcolor=blue]{hyperref}

\begin{document}

\title{NeuralCMS: A deep learning approach to study Jupiter's interior}
\author{M. Ziv\inst{1}, E. Galanti\inst{1}, A. Sheffer\inst{1}, S. Howard\inst{2,3}, T. Guillot\inst{2},\and Y. Kaspi\inst{1}}
\institute{Department of Earth and Planetary Sciences, Weizmann Institute of Science, Rehovot 76100, Israel\\
\email{maayan.ziv@weizmann.ac.il}\and Université Côte d'Azur, Observatoire de la Côte d'Azur, CNRS, Laboratoire Lagrange, France\and Institut für Astrophysik, Universität Zürich, Winterthurerstr. 190, 8057 Zürich, Switzerland}
\date{Received 3 April 2024 / Accepted 6 May 2024}
\abstract 
{NASA's Juno mission provided exquisite measurements of Jupiter's gravity field that together with the Galileo entry probe atmospheric measurements constrains the interior structure of the giant planet. Inferring its interior structure range remains a challenging inverse problem requiring a computationally intensive search of combinations of various planetary properties, such as the cloud-level temperature, composition, and core features, requiring the computation of $\sim$\(10^9\) interior models.}
{We propose an efficient deep neural network (DNN) model to generate high-precision wide-ranged interior models based on the very accurate but computationally demanding concentric MacLaurin spheroid (CMS) method.}
{We trained a sharing-based DNN with a large set of CMS results for a four-layer interior model of Jupiter, including a dilute core, to accurately predict the gravity moments and mass, given a combination of interior features. We evaluated the performance of the trained DNN (NeuralCMS) to inspect its predictive limitations.}
{NeuralCMS shows very good performance in predicting the gravity moments, with errors comparable with the uncertainty due to differential rotation, and a very accurate mass prediction. This allowed us to perform a broad parameter space search by computing only $\sim$\(10^4\) actual CMS interior models, resulting in a large sample of plausible interior structures, and reducing the computation time by a factor of \(10^5\). Moreover, we used a DNN explainability algorithm to analyze the impact of the parameters setting the interior model on the predicted observables, providing information on their nonlinear relation.}
{}
\titlerunning{}
\authorrunning{Ziv, M., et al.}
\maketitle

\section{Introduction}

\begin{figure*}
\sidecaption
    \includegraphics[width=12cm]{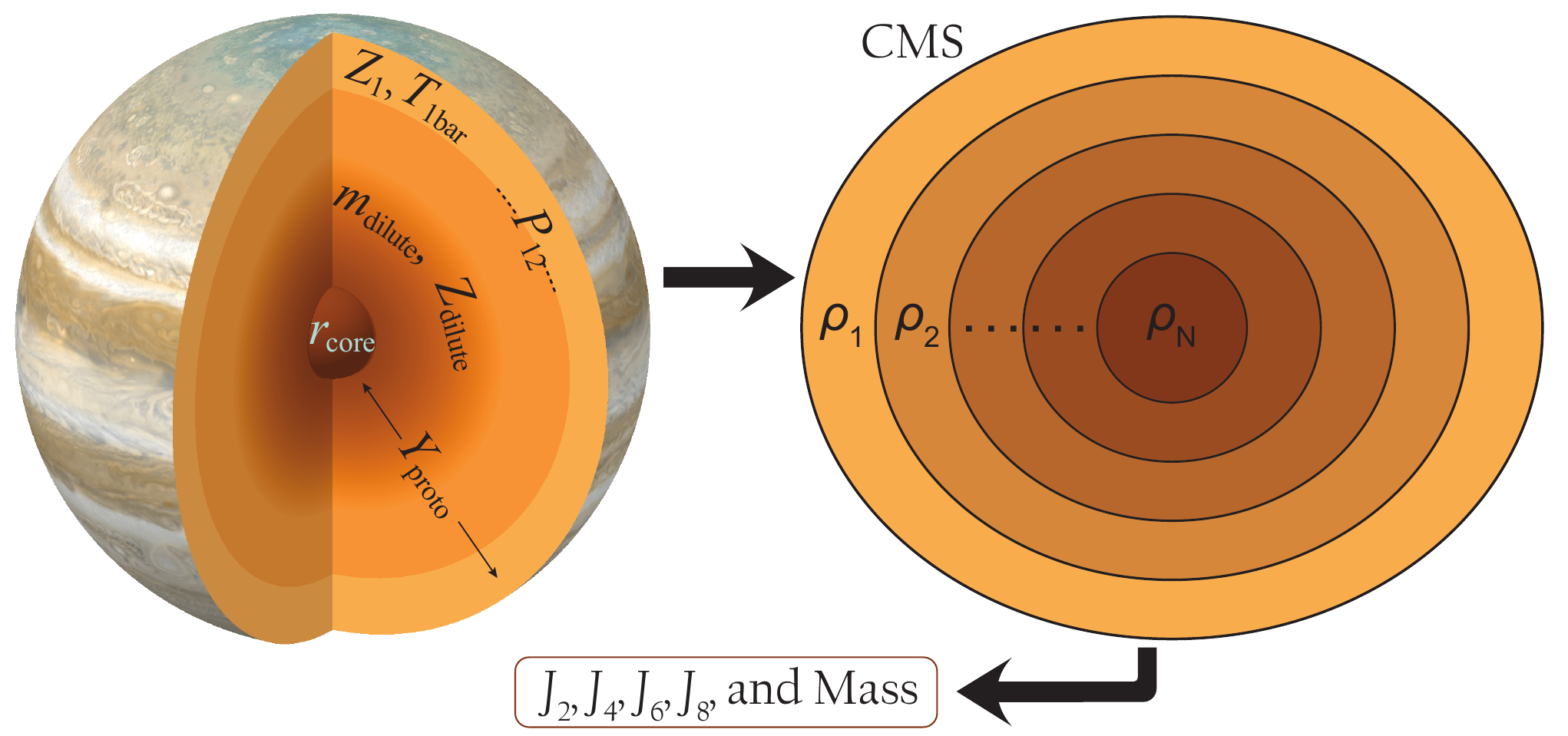}
    \caption{Schematic view of Jupiter's dilute core model used in this study and the computational process: given a combination of the seven marked interior parameters (left), the CMS method (right) converges to solve the gravity moments and mass, is then compared to the Juno measurements to determine the feasibility. The image in Jupiter's schematic (left) is available at \url{https://www.planetary.org/space-images/merged-cassini-and-juno}.}
    \label{fig: Jupiter_scheme}
\end{figure*}

The interior structure of Jupiter holds information on its formation and evolution processes, with the two research fields being highly related to one another \citep{Helled2022, Miguel2022}. The range of plausible interior structures of Jupiter is constrained by the accurately measured gravity field by  NASA's Juno mission \citep{Bolton2017, Iess2018, Durante2020} and atmospheric measurements by both Juno \citep{Li2020} and the Galileo probe \citep{vonZahn1998, Seiff1998, Wong2004}. In addition, it is also affected by the surface winds and their internal structure, which significantly contribute to the gravity field \citep{Kaspi2018, Kaspi2023}. Inferring this range requires the exploration of a large parameter space of interior models to identify those consistent with the observations.

Relating the above observables to the physical parameters defining the interior structure of a gas giant can be done by two approaches. Theory of figures (ToF) \citep{Zharkov1978}, implemented for example to the seventh order in \cite{Nettelmann2021} and to the fourth order \citep{Nettelmann2017} in the CEPAM model \citep{Guillot1995b, Guillot2018}, which was used by \cite{Howard2023a}, and the more accurate concentric Maclaurin spheroid (CMS) method \citep{Hubbard2013}, used by \cite{Militzer2022}, which is more computationally demanding \citep{Militzer2019}. One way to overcome the computational burden of the CMS approach is to correct the ToF results with offsets to the gravity moments to make up for the precision difference \citep{Guillot2018, Miguel2022}. However, the offsets are defined for specific parameters and might not represent the entire parameter space.

Previous studies have suggested deep learning approaches to characterize the interior of exoplanets by predicting the distribution of interior features given the planetary mass, radius, and several additional parameters (e.g., the fluid Love number \(k_2\), the effective temperature, the temperature at one bar), thus addressing the inverse problem directly \citep{Baumeister2020, Zhao2022, Baumeister2023}. Recently, \cite{Haldemann2023} presented an approach allowing inference of both the inverse problem and the forward interior model.

In this work, we present NeuralCMS, a new approach to accelerate the CMS method, by predicting the model results using a deep neural network (DNN), which in practice can quickly compute millions of interior models simultaneously, or a single interior model on the order of milliseconds. Theoretically, DNNs are a suitable choice to regress the CMS results as they can approximate any nonlinear function between an adjustable number of inputs and outputs \citep{Goodfellow2016}. We used a DNN for the principal task of approximating the detailed forward CMS model constrained by the gravity moments and mass. NeuralCMS can then be used in any search algorithm, such as Monte Carlo methods \citep{Miguel2022, Militzer2023b}, to assemble a sample of plausible interior structures. We also demonstrate that with the advance in explainable DNN techniques \citep{Samek2021}, further investigation of the nonlinear relations between interior features and the observables can be made possible.

In Sect. \ref{sect2} we describe the numerical and theoretical interior model, followed by a description of the dataset used to train the DNN. In Sect. \ref{sect3} we present the DNN architecture, performance, and training specifics. In Sect. \ref{sect4} we present the efficiency derived from our approach by performing a simple grid search for plausible Jupiter's interior solutions.

\section{Jupiter interior structure model}\label{sect2}

Our numerical interior model is based on a publicly available CMS model \citep{MovshovitzCMS, Movshovitz2020}. CMS \citep{Hubbard2013} is an iterative method to compute the shape and gravity harmonics (\(J_{2n}\)) of a rotating fluid planet. It is constructed of multiple concentric Maclaurin spheroids set by their equatorial radii, and using the hydrostatic equilibrium of the gravitational and rotational potential, it solves the shape for each spheroid assembling the planet (Fig. \ref{fig: Jupiter_scheme}). We modeled Jupiter with \(N=1041\) spheroids spaced the same as in \cite{Howard2023a}. We validated our CMS model against the analytic \(n=1\) polytrope solution \citep{Wisdom2016} (see Appendix \ref{apendix:CMS validation}).

We constructed a four-layer model of Jupiter as shown in Fig. \ref{fig: Jupiter_scheme}, similar to \cite{Miguel2022} and \cite{Howard2023a}. The outer layer is mostly composed of hydrogen and helium with their mass fraction \(X_1\) and \(Y_1\), respectively. We set \(Y_{1}/(X_{1}+Y_{1})=0.238\) to be consistent with the Galileo probe measurements \citep{vonZahn1998}. The mass fraction of heavier elements, or metallicity, in this layer, is marked by \(Z_1\), which was constrained in the atmosphere by both Juno and Galileo to be higher than the solar abundance \citep{Wong2004, Li2020, Howard2023a}. Recent interior models still struggle to reconcile with this important observation \citep{Howard2023c}. The outer envelope is treated as adiabatic, with a constant entropy determined by the temperature at one bar, which was measured by the Galileo probe to be \(T_{\rm{1bar}}=166.1\pm0.8\,\rm{K}\) \citep{Seiff1998}, and was recently suggested to reach \(T_{\rm{1bar}}=170.3\pm3.8\,\rm{K}\) after reassessing Voyager radio occultations \citep{Gupta2022}.

The boundary between the inner He-rich and the outer He-poor envelopes is set by the pressure \(P_{12}\) representing a region where immiscibility of He in H occurs, and based on simulations of phase separation of H and He mixtures should occur between $\sim$0.8 and $\sim$3 Mbar \citep{Morales2013}. We set the metallicity of the inner envelope to be \(Z_2=Z_1\). Then we implemented a dilute core by imposing an inward increase in the mass fraction of heavy elements using the same formulation used by \cite{Miguel2022} with two main controlling parameters, \(Z_{\rm dilute}\) defining the maximum mass fraction of heavy elements in the dilute core, and \(m_{\rm dilute}\) representing the extent of the dilute core in normalized mass (see Appendix \ref{apendix: training data} for more details). The helium mass fraction in the inner envelope and the dilute core regions is forced by requiring the planet's overall helium abundance to be consistent with the protosolar value, \(Y_{\rm{proto}}=0.278\pm 0.006\) \citep{Serenelli2010}. Most recent models agree on the presence of a dilute core inside Jupiter although its extent exhibits discrepancies between interior and formation models. Recently, models with a small enough dilute core consistent with Juno were suggested \citep{Howard2023a}. Finally, we allowed the presence of a compact core composed of heavy materials only, and its normalized radius \(r_{\rm{core}}\) controls it.

It was shown that the choice of the equation of state (EOS) for hydrogen and helium strongly affects the interior model \citep{Miguel2016, Howard2023a}. For this work, we did not explore this effect but used only the pure H and He tables from \cite{Chabarier2019}, and the nonideal mixing effect tables, accounting for the interactions between H and He, from \cite{Howard2023b}. We used the Sesame water EOS \citep{Lyon1992} for heavy materials. We used the additive volume law combined with the nonideal corrections to compute the density and entropy at a given pressure and temperature \citep{Howard2023b, Howard2023c}. The EOS for the compact core is the analytical solution from \cite{Hubbard1989}.

Unlike other CMS-based models, we did not restrict the calculated planetary mass to a specific value. As we treat the gravity moments, the mass is compared to the Juno-derived value within its uncertainty. The mass was computed through the observed \(GM=1.266865341\times10^{17}\,\rm{m^3\,s^{-2}}\) \citep{Durante2020}, and while in our model we did not use the Newtonian constant of gravitation \(G\) explicitly, the range of its suggested values results in a noticeable uncertainty in Jupiter's mass. Dividing the measured \(GM\) by the extremum values of \(G\) collected by CODATA \citep{Tiesinga2021} gives a mass range between 1.8978 and 1.8988 \(10^{27}\,\rm{kg}\). We used \(G=6.673848\times10^{-11}\rm{m^3\,kg^{-1}\,s^{-2}}\) \citep{Mohr2012}, to be consistent with \cite{Howard2023a}, resulting in \(M_{\rm{J}}=1.8983\pm0.0005\times10^{27}\,\rm{kg}\). Also, deviations in the equatorial radius \(R_{\rm{eq}}\), stemming from either measurement errors \citep{Lindal1992} or from the dynamical height due to the wind \citep{Galanti2023}, suggest a similar mass uncertainty.

We reduced the characterization of Jupiter's interior to the seven input parameters shown in Fig. \ref{fig: Jupiter_scheme} and allowed them to vary, while \(R_{\rm{eq}}=71\,492\,\rm{km}\) \citep{Lindal1992} and Jupiter's rotation rate of 9 hr 55 min 29.7s \citep{Riddle1976} were kept constant. For training, we used previously computed results of over \(10^6\) CMS interior models, with different setups of 7D input samples: (1) sparsely gridded inputs; (2) randomly sampled inputs, both resulting in a large range in the outputs \(J_{2n}\) and mass; and (3) densely gridded inputs where the outputs are closer to the Juno measurements. The deep learning model was trained to accurately predict a broad range of interior models, to be used in a wide search for plausible models. The training dataset range and distribution are presented in Table \ref{tab:training_range} and in Fig. \ref{fig: Training_dist}.

\section{A deep sharing-based neural network}\label{sect3}

\begin{figure}
    \centering
    \includegraphics[width=1\linewidth]{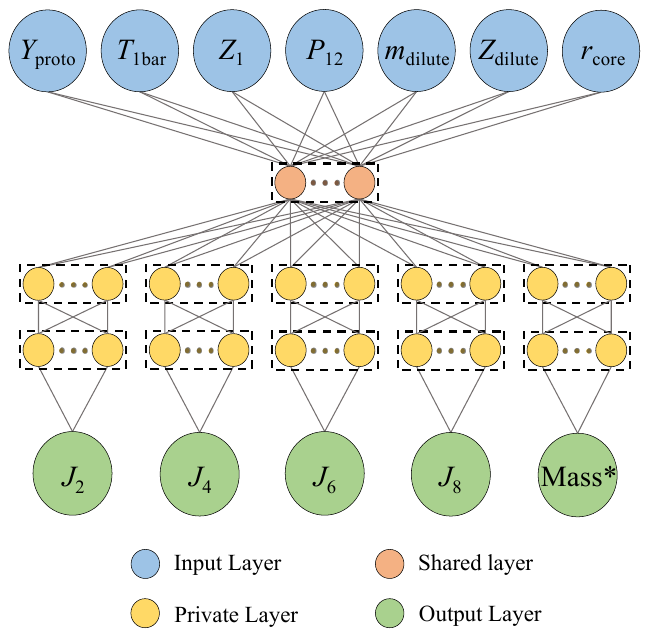}
    \caption{Schematic diagram of the DNN architecture presented in this study. The hidden layers, both shared and private are marked by dashed black outlines and contain 1024 neurons each. The mass is used for training on the gravity moments \(J_{2n}\) and it is predicted separately.}
    \label{fig:NN_scheme}
\end{figure}

In recent years, feedforward neural networks have been a popular machine-learning approach in many research fields. They are capable of deciphering information and relations in multidimensional data \citep{LeCun2015}. They are generally composed of input and output layers connected through several hidden layers, all built from a varying number of neurons. Their training is controlled by minimizing a so-called loss function between the predicted and the true output values, practically by optimizing the weights and biases on the connections between neurons \citep{LeCun2015}. Many machine-learning algorithms were suggested to address multi-target regression problems where the outputs are correlated \citep{Borchani2015, Cui2018}.

For this work, we adopted a sharing-based architecture \citep{Caruana2002,Reyes2019}, shown in Fig. \ref{fig:NN_scheme}. Our feedforward DNN comprises an input layer fully connected to a shared hidden layer with 1024 neurons. The shared layer is fully connected to five separate private hidden blocks, one for each output. Each block contains two 1024-neuron fully connected layers computing a single output value. The internal hidden layers are activated using the nonlinear Rectified Linear Unit (ReLU) function \citep{Goodfellow2016}, which is commonly used and easy to optimize. Since the gravity moments and mass are all functions of the density structure and shape of the planet, the correlation between them is designed to be learned by the shared layer. The private layers then act as a single output regressor. We find that using the mass as an output parameter improves the prediction of \(J_{2n}\), but it is more precisely predicted using an entirely separate DNN that has the same seven input parameters, and four fully connected layers with 1024, 512, 256, and 128 neurons, respectively, each activated with ReLU, predicting only the mass. Appendix \ref{apendix: architectures}  discusses other architectures that were tested.

\begin{figure}
    \centering
    \includegraphics[width=1\linewidth]{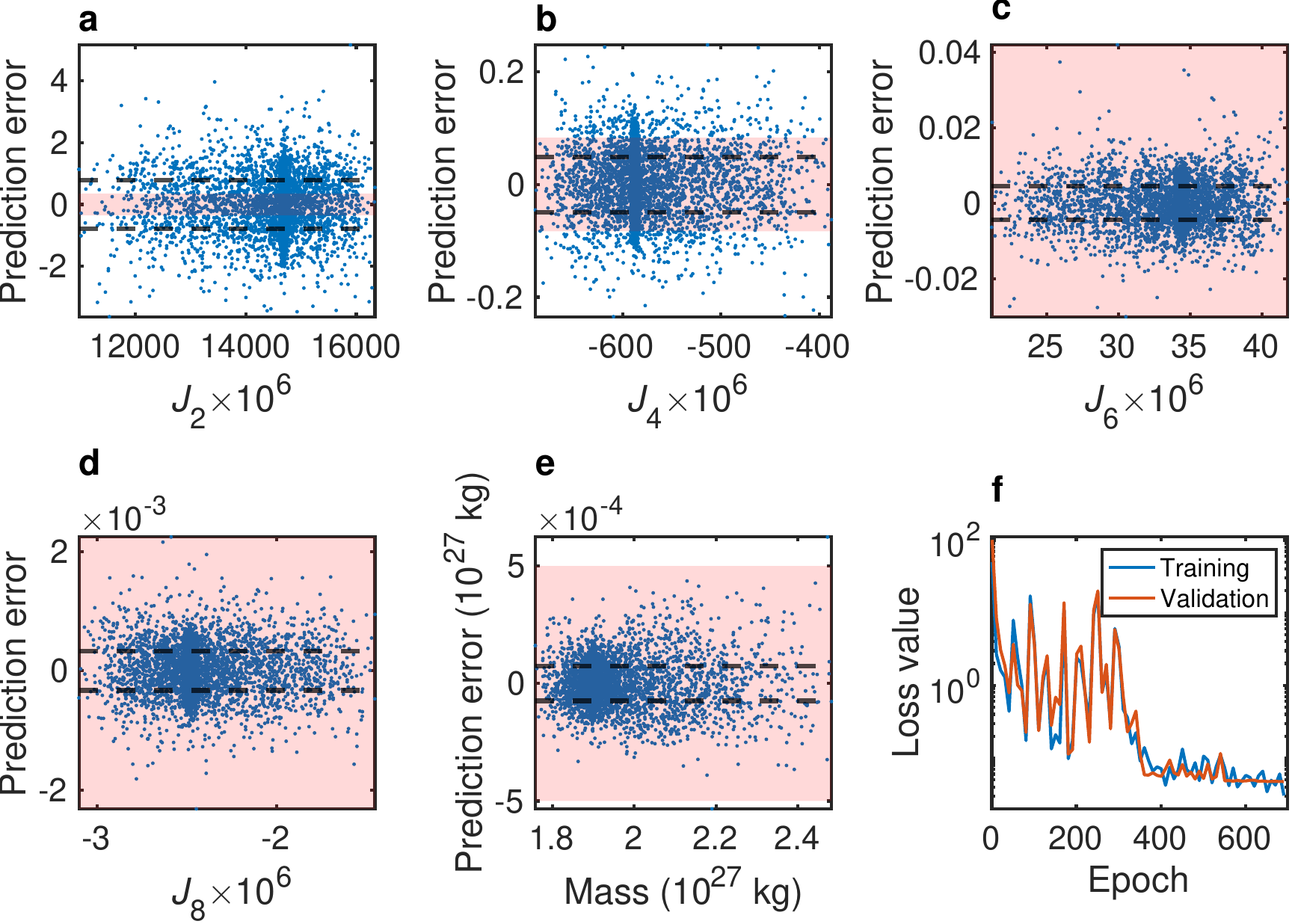}
    \caption{Performance of NeuralCMS on a sample of \(10^4\) models from the validation dataset (\textbf{a-e}). The dashed black lines are the standard deviation of the full validation dataset error \(\epsilon_{\sigma}\). The red patch represents the combined uncertainty from dynamics \citep{Miguel2022} and measurement errors \citep{Durante2020} for the gravity harmonics: \(\sqrt{(3\sigma_{\mathrm{wind}})^2+(3\sigma_{\mathrm{Juno}})^2}\). The mass uncertainty is due to various \(G\) values. The learning curve of the DNN at every ten epochs using Eq. \ref{eq:loss} (\textbf{f}).}
    \label{fig: accuarcies}
\end{figure}

We normalized the inputs to have values between zero and one, using the bounds of the training dataset (see Table \ref{tab:training_range}). The gravity moments were scaled by \(10^6\) and taken in positive values. The mass was scaled by \(10^{-27}\). We initialized the DNN weights using Kaiming uniform distribution to prevent activation from potentially harming the training \citep{He2015}. During training, we approached the predicted and true output values using a weighted mean squared error loss function,
\begin{equation}
L=\frac{1}{NW}\sum_{i=1}^{N}\left(\left(\frac{M_{i}^{{\rm pred}}-M_{i}^{{\rm CMS}}}{\Delta M}\right)^{2}+\sum_{n=1}^{4}\left(\frac{J_{2n,i}^{{\rm pred}}-J_{2n,i}^{{\rm CMS}}}{3\sigma_{2n}}\right)^{2}\right)
\label{eq:loss}
,\end{equation}
where \(N\) is the number of samples, \(\Delta M=0.0005\times10^{27}\,\rm{kg}\) is the mass uncertainty discussed in Sect. \ref{sect2}, \(3\sigma_{2n}\) is Juno's \(3\sigma\) uncertainty for \(J_{2n}\) \citep{Durante2020}, and \(W=\frac{1}{(\Delta M)^2}+\sum_{n=1}^{4}\frac{1}{(3\sigma_{2n})^2}\) is the sum of the weights. This gives a larger weight to the more accurately measured observables. The loss function was minimized using the Adam optimizer \citep{Kingma2014} with a learning rate starting from 0.001 and reduced tenfold at manually chosen epochs. We took 80\% of the models from the full dataset for training, leaving the rest to validate the trained DNN. The DNN was trained for 700 epochs using the PyTorch library \citep{Paszke2019}, showing no overfitting (Fig. \ref{fig: accuarcies}f).

The performance of our DNN was evaluated by the prediction errors \(\epsilon\) for each output. Figure \ref{fig: accuarcies} shows the prediction errors as a function of the true output values compared to the uncertainty stemming from measurement errors \citep{Durante2020} and the wind \citep{Galanti2023, Miguel2022}. For all model outputs, except for \(J_2\), the standard deviation of the prediction error \(\epsilon_{\sigma}\) is smaller than the combined uncertainty, and it is mostly comparable to the wind-derived uncertainty. Specifically, for \(J_{2}\times10^6\), \(\epsilon_{\sigma}=0.789\) is about twice the wind-related uncertainty but smaller than the offset applied to ToF results by a factor of $\sim$2-7 \citep{Guillot2018, Miguel2022}. \cite{Debras2018} evaluated the uncertainty on \(J_2\times10^6\) related to assumptions of the CMS method to be roughly 0.1, being lower than \(\epsilon_{\sigma}\). Deviations in \(J_{2}\times10^6\) from the analytical solution for the \(n=1\) polytrope found in previous studies with a similar number of spheroids are between $\sim$0.1-3 \citep{Wisdom2016, Guillot2018, Debras2018, Nettelmann2021}. We note that the ToF offsets and deviations from the polytropic solutions are systematic errors, whereas \(\epsilon_{\sigma}\) are both positive and negative (see Fig. \ref{fig: accuarcies}). Table \ref{tab:perfomance} compares these sources for errors and uncertainty with the DNN's performance. 

The relatively small prediction errors allow the DNN to eliminate the vast majority of interior models that deviate from the Juno measurements. Moreover, the prediction errors are independent of the output values, highlighting the DNN's ability to predict a large range of interior models. Due to the very accurate measured \(J_2\) and \(J_4\) compared to the DNN errors, interior models still need to be calculated using CMS, to eliminate models falsely predicted by the DNN to be consistent with the observations. Also, the CMS output contains valuable information we wish to retrieve such as density and composition profiles.

\section{Performance and interpretation of NeuralCMS}\label{sect4}

\begin{figure}
    \centering
    \includegraphics[width=1\linewidth]{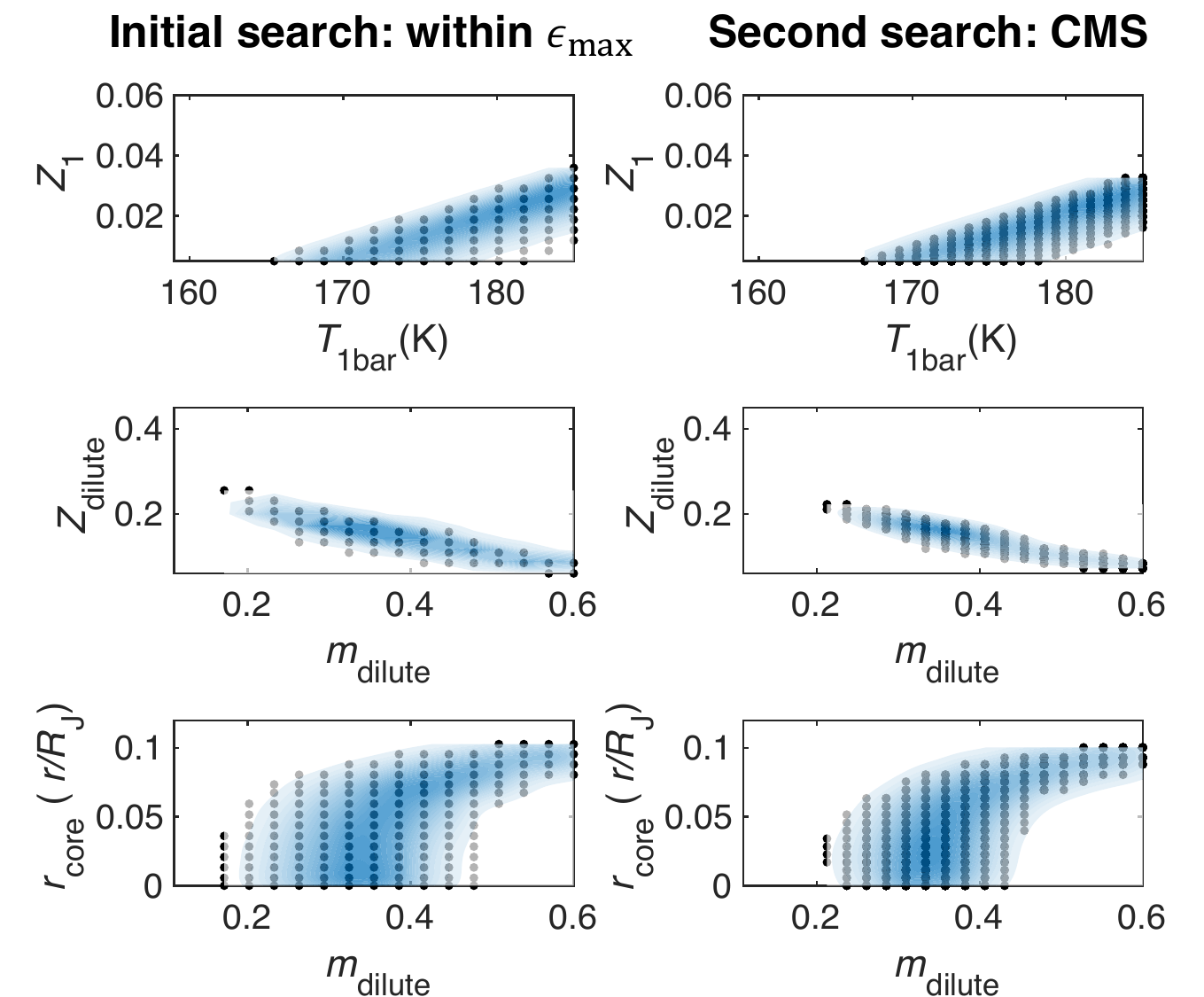}
    \caption{Correlation between interior features for the two grid search stages. The black points are model results, and the blue shading is the models' distribution. The left column shows accepted models predicted by NeuralCMS in the first grid search, within the DNN's maximal absolute prediction errors on the validation dataset. In the right column are accepted models computed with CMS found in the second tighter grid search. The axes range is the initial wide search range. The range of \(P_{12}\) and \(Y_{\rm proto}\) was not reduced. The middle panels nicely reproduce previous results \citep{Howard2023a}.}
    \label{fig:grid_search}
\end{figure}

To demonstrate the computational efficiency gained by using NeuralCMS, we performed a simple grid search exploring all possible combinations of an equally spaced grid for each of the seven parameters with \(m\) grid points. We initiated the first grid search using only the DNN, with a wide range for all the parameters using the bounds shown in the axes range in Fig. \ref{fig:grid_search}, the determined \(Y_{\rm{proto}}=0.278\pm 0.006\) \citep{Serenelli2010}, and \(P_{12}\) between 0.8 and 5 Mbar,  with \(m=17\) grid points for each parameter, exploring over \(4\times10^8\) interior models. This procedure takes $\sim$2 hours. The results of this grid search were used to reduce the range of input parameters by eliminating models that do not fall within a wind-effect criterion regarding the Juno measurements within the absolute maximal prediction errors on the validation dataset, \(\epsilon_{\rm max}\). The wind-effect criterion is a range we allowed models to deviate from the Juno measurements setting a subsequent exploration of the effects of a coupled wind model. The criterion accepts models that are within \(2\times10^{-6}\) and \(10^{-6}\) from the measured \(J_2\) and  \(J_4\), respectively, and within the mass uncertainty discussed in Sect. \ref{sect2}. These values were added to the prediction errors considered. The range of \(Z_1\) and \(Z_{\rm dilute}\) was significantly reduced after the first grid search as shown in the left column of Fig. \ref{fig:grid_search}.

Using NeuralCMS, we encountered the known difficulty of finding solutions that are consistent with the Juno gravity measurement, the Galileo-measured \(T_{\rm 1bar}\), and the high measured atmospheric metallicity \citep{Howard2023c}. Moreover, the correlation plot between \(m_{\rm dilute}\) and \(Z_{\rm dilute}\) shown in the middle-left panel of Fig. \ref{fig:grid_search} is similar to the results produced with the same EOS by \cite{Howard2023a} and with a slightly different setup (see their Fig. 13). This provides another validation for our model. We note that some of the accepted models here, shown in the left column of Fig. \ref{fig:grid_search}, will be eliminated with CMS calculations because we considered large prediction errors due to the relatively sparse grid used, hence only serving to reduce the interior parameters' range roughly.

The second grid search was done with the narrowed parameter range obtained from the initial grid search, with a denser grid of \(m=20\) grid points per parameter, exploring \(1.28\times 10^9\) interior models. Again, we used the wind-effect criterion, now added to the \(3\sigma\) prediction errors on the validation dataset (\(\epsilon_{3\sigma}\)) to retrieve 10871 possible models to test against actual CMS calculations. From these possible combinations, we find 2927 interior models accepted by the wind-effect criterion according to their CMS results. The initial search was enough to produce a compact distribution of models with respect to the parameters presented in Fig. \ref{fig:grid_search}, which was further tuned after the second denser grid search. Testing the grid search predictions made by the DNN against the CMS outputs yields a performance similar to that observed on the validation dataset. Importantly, using NeuralCMS, only $\sim$\(10^4\) actual CMS models need to be computed instead of the impractical computation of $\sim$\(10^9\) CMS models, to assemble a sample of $\sim$3000 plausible interior models, all falling within \(\epsilon_{3\sigma}\).

\begin{figure}
    \centering
    \includegraphics[width=1\linewidth]{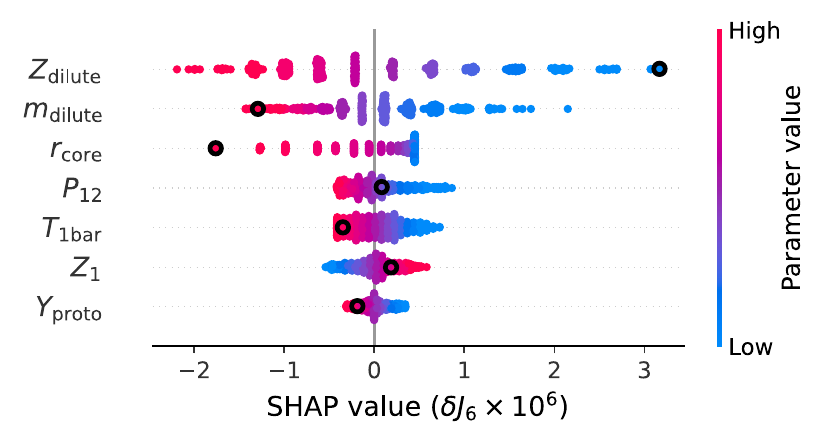}
    \caption{Contribution in ppm to the prediction of \(J_6\) of 500 interior models (each point in a row is an individual model) that are consistent with the Juno measurements, within the wind-effect criterion. Higher SHAP values correspond to a larger contribution to the predicted \(J_6\). Points are stacked vertically where there is a high density of model solutions. The colors scale the values of each interior parameter. For example, when high, only \(Z_1\) positively contributes to \(J_6\). The black circles correspond to a specific interior model having the highest SHAP value for \(Z_{\rm dilute}\).}
    \label{fig:shap}
\end{figure}

The DNN can also be used to reveal the contribution of the interior parameters to the predictions. The SHapley Additive exPlanations (SHAP) values are a game theory approach assigning an impact value for each model input representing its contribution to a specific prediction \citep{Lundberg2017}. SHAP values provide a locally accurate approximation such that for a single prediction, the SHAP values for all inputs sum to the deviation of the predicted values from their mean in a reference dataset. This provides an interpretation of the magnitude and direction in which each input moves the model predictions. For this work, we used Deep SHAP \citep{Lundberg2017}, which linearizes the DNN's nonlinear components to backpropagate the SHAP computation through the network. We took all CMS-accepted models as the reference dataset and calculated SHAP values for 500 random models from these accepted models. As an example, we examined the SHAP values for \(J_6\), an observable that is usually difficult to fit \citep{Debras2019, Militzer2022}. Figure \ref{fig:shap} shows these SHAP values for all 500 models, each corresponding to a point on each row. For example, we marked with black circles a specific model with the highest SHAP value for \(Z_{\rm dilute}\) (see Table \ref{tab:shap}). The parameters controlling the planet's core (\(m_{\rm dilute}\), \(Z_{\rm dilute}\), and \(r_{\rm{core}}\)) have the largest effect on the prediction of \(J_6\). Moreover, the analysis shows that all model parameters have an overall monotonic effect on \(J_6\) (color-coded in Fig. \ref{fig:shap}). The analysis also highlights the interplay among input features. The SHAP values for \(T_{\rm{1bar}}\) and \(Z_1\) exhibit similar magnitudes but in opposite directions, such that high values of \(Z_1\) (\(T_{\rm{1bar}}\)) contribute positively (negatively), effectively balancing out each other's contribution to \(J_6\). Conversely, the dilute core parameters yield the same sign contribution to \(J_6\). Similar results are observed for the other gravity harmonics.

\section{Conclusion}\label{conclusion}

We present NeuralCMS, an efficient deep learning approach to explore the range of plausible interior structures of Jupiter, constrained by the Juno-measured gravity field and mass. This is done by training a DNN to predict the results of the sophisticated and computationally demanding CMS method. We trained a sharing-based DNN using results from over \(10^6\) CMS calculations, showing good performance compared to the uncertainties associated with the gravity moments and mass. 

We show that NeuralCMS can be used to eliminate interior models inconsistent with the measured gravity field and substantially reduce the number of actual CMS runs. We demonstrate the efficiency of NeuralCMS by performing a grid search for model solutions consistent with Juno. Evaluating over \(10^9\) possible models with NeuralCMS allowed us to identify $\sim$\(10^4\) solutions on which actual CMS runs were performed, producing a big sample of nearly 3000 plausible interior models, thus reducing the computational time by a factor of \(10^5\). This would not have been computationally feasible using only the CMS model. 

We demonstrate that despite the DNN's complex nature, it is possible to interpret relations between the physical interior parameters and their contribution to the observables using SHAP values. As an example, we show that within their range relevant to the Juno measurements, parameters controlling the planet's core have the largest impact on the predicted \(J_6\) suggesting that high dilute core metallicity is associated with small dilute core extent, and vice versa, thus compensating for each other.

NeuralCMS can be used in any search methodology to detect plausible interior structures without a single CMS computation, acknowledging its prediction errors. It can also be further expanded to allow additional interior parameters (e.g., the equatorial radius and a temperature jump in the He rain region) and to more complex interior structures of Jupiter or other gaseous planets. NeuralCMS is available on GitHub\footnote{https://github.com/zivmaaya/NeuralCMS}.

\begin{acknowledgements}
This work was supported by the Israeli Space Agency and the Helen Kimmel Center for Planetary Science at the Weizmann Institute.
\end{acknowledgements}

\bibliographystyle{aa}
\bibliography{manuscript}

\begin{appendix}
\clearpage 
\section{CMS model validation}\label{apendix:CMS validation}

To validate our CMS model, we tested it against the analytic Bessel solution for Jupiter's uniformly rotating \(n=1\) polytrope \citep{Wisdom2016}. This was done using the same spheroid radii grid as used by \cite{Howard2023a}. Table \ref{tab:index1} shows our model convergence to the analytic Bessel solution with an increasing number of spheroids. Our model exhibits good convergence with \(N=1024\) spheroids.

\begin{table*}
    \caption{Validation of our CMS model for Jupiter's polytropic of index unity model with an increasing number of CMS spheroids $(N)$.}
    \label{tab:index1}
    \centering
    \begin{tabular}{l c c c c c}
    \hline\hline
     & \(J_{2}\times10^{2}\)\ & \(-J_{4}\times10^{4}\) & \(J_{6}\times10^{5}\) & \(-J_{8}\times10^{6}\)  & \(J_{10}\times10^{7}\) \\
    \hline
        WH16 CMS $(N=512)$   & 1.3989240 & 5.3187921 & 3.0122304 & 2.1324587 & 1.7409889\\
        This work $(N=512)$  & 1.3990044 & 5.3193288 & 3.0126663 & 2.1328713 & 1.7413831\\
        This work $(N=1024)$ & 1.3989138 & 5.3187281 & 3.0121981 & 2.1324535 & 1.7409727\\
        This work $(N=2048)$ & 1.3988791 & 5.3184861 & 3.0120029 & 2.1322751 & 1.7407947\\
        This work $(N=4096)$ & 1.3988644 & 5.3183798 & 3.0119152 & 2.1321937 & 1.7407126\\
        Bessel             & 1.3988511 & 5.3182810 & 3.0118323 & 2.1321157 & 1.7406712\\
    \hline
    \end{tabular}
    \tablefoot{The Bessel solution is taken from \cite{Wisdom2016} (WH16).}
\end{table*}

\section{Dilute core formulation and the training data range}\label{apendix: training data}

Jupiter's dilute core is implemented differently in various interior models. For this Letter, we followed the formulation of \cite{Miguel2022}, defining the mass fraction of heavy elements in the inner envelope and the dilute core,
\[\Vec{Z}=Z_{2}+\frac{Z_{\mathrm{dilute}}-Z_{2}}{2}\left[1-\mathrm{erf}\left(\frac{\Vec{m}-m_{\mathrm{dilute}}}{\delta m_{\mathrm{dil}}}\right)\right],\]
where \(Z_{\rm dilute}\) is the maximum metallicity in the dilute core, \(m_{\rm dilute}\) represents the extent of the dilute core in normalized mass, and \(\delta m_{\rm dil}\) controls the steepness of the metallicity gradient and was set to be \(\delta m_{\rm dil}=0.075\). This region is not an adiabat, but we treat each spheroid as an adiabat. We infer the interpolated entropy (\(S\)) and temperature (\(T\)) profiles, such that for each spheroid \(i\) in the dilute core region, \(T_{i+1}=T(P_{i+1},S_{i},Y_{i},Z_{i})\) and \(S_{i+1}=S(P_{i+1},T_{i+1},Y_{i+1},Z_{i+1})\).

\begin{figure*}
    \centering
    \includegraphics[width=1\linewidth]{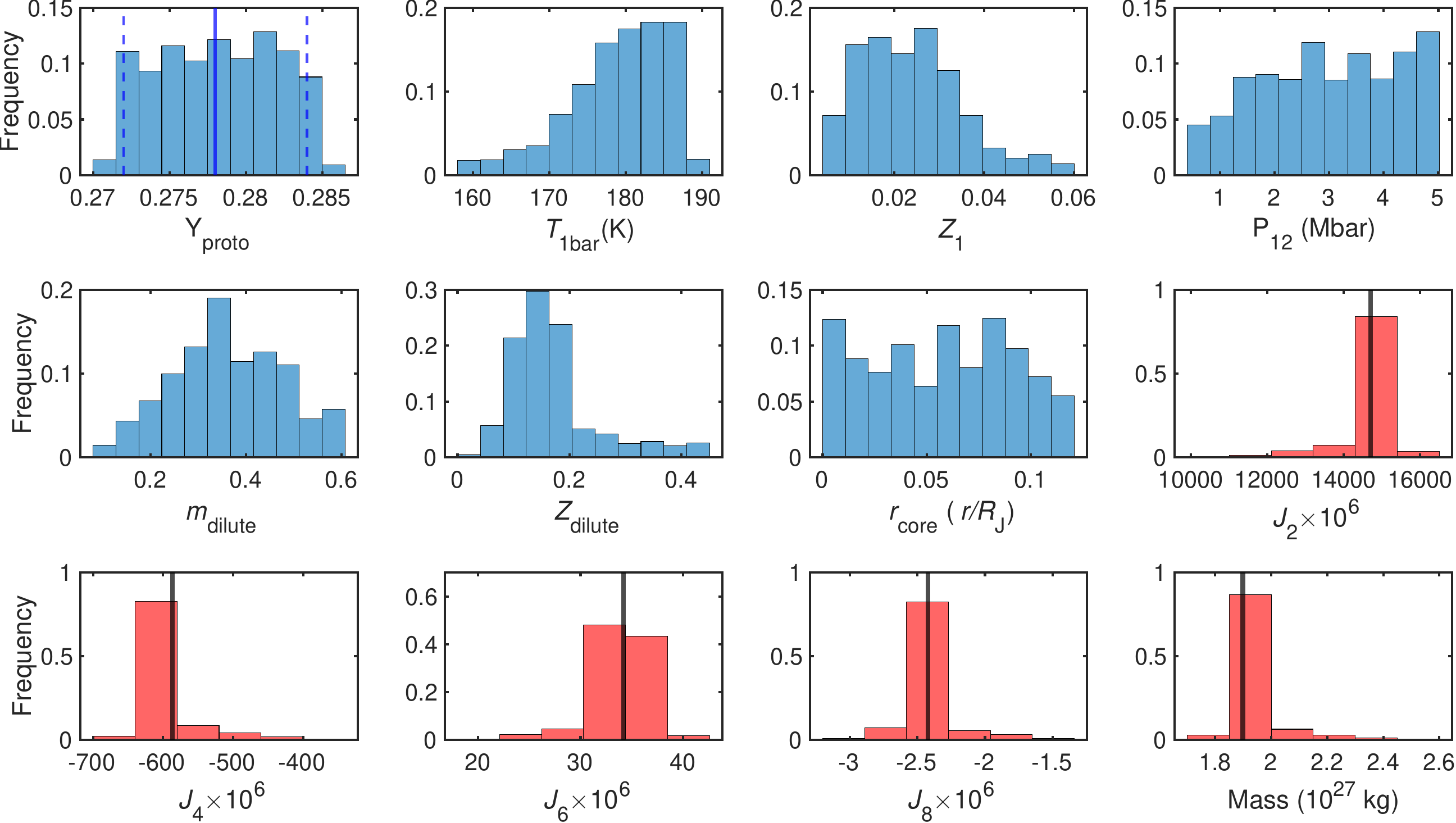}
    \caption{Distribution of the inputs (blue) and the outputs (red) in the training dataset. For \(Y_{\rm{proto}}\) the solid blue line is the determined value and the dashed blue lines represent the uncertainty \citep{Serenelli2010}. The solid black lines are Juno-derived values \citep{Durante2020}. We note that some of the histogram bins for the outputs are not visible due to the high frequency of the proximate to Juno models. We refer readers to Table \ref{tab:training_range} and the x-axis range for the full range of the dataset.}
    \label{fig: Training_dist}
\end{figure*}

Table \ref{tab:training_range} and Fig. \ref{fig: Training_dist} show the parameter range and distribution of the training and validation datasets. The range of most parameters is larger than observational or theoretical constraints to allow for a broad exploration of the parameter space.

\begin{table*}
    \caption{Parameter range of the training and validation datasets.}
    \label{tab:training_range}
    \centering
    \begin{tabular}{l c c c}
    \hline\hline
     Parameter & Lower value & Upper value & Observation \\
    \hline
        \(Y_{\rm{proto}}\)   & 0.270 & 0.286 & \(0.278\pm 0.006\)\tablefootmark{a}\\
        \(T_{\rm{1bar}}\,(\rm{K})\)  & 159 & 190 & 166.1$\pm$0.8\tablefootmark{b}; 170.3$\pm$3.8\tablefootmark{c}\\
        \(Z_1\) & 0.005 & 0.06 & \(\approx0.046=3Z_{\mathrm{protosolar}}\)\tablefootmark{d} \\
        \(P_{12}\,(\rm{Mbar}\)) & 0.5 & 5 \\
        \(m_{\rm dilute}\) & 0.11 & 0.6 \\
        \(Z_{\rm dilute}\) & 0.01 & 0.45 \\
        \(r_{\rm{core}}\,(r/R_{\rm{J}})\) & 0 & 0.12 \\
    \hline
        \(J_2\times 10^{6}\) & 10632 & 16452 & 14696.5735$\pm$0.0017\tablefootmark{e} \\
        \(J_4\times 10^{6}\) & $-696.5$ & $-369.5$ & $-586.6085\pm0.0024$\tablefootmark{e} \\
        \(J_6\times 10^{6}\) & 19.92 & 42.37 & 34.2007$\pm$0.0067\tablefootmark{e} \\
        \(J_8\times 10^{6}\) & $-3.151$ & $-1.356$ & $-2.422\pm0.021$\tablefootmark{e} \\
        Mass\,(\(10^{27}\,\rm{kg}\)) & 1.732 & 2.590 & 1.8983$\pm$0.0005\tablefootmark{f} \\
    \hline
    \end{tabular}
    \tablefoot{In all models we kept \(Z_1\leq Z_{\rm dilute}\). The lower part of the table shows the model outputs.\\
    \tablefoottext{a}{\cite{Serenelli2010}, }
    \tablefoottext{b}{\cite{Seiff1998}, }
    \tablefoottext{c}{\cite{Gupta2022}, }
    \tablefoottext{d}{\cite{Howard2023a}, }
    \tablefoottext{e}{\cite{Durante2020}, }
    \tablefoottext{f}{see Sect. \ref{sect2}.}
    }
\end{table*}

\section{Exploration of DNN architectures}\label{apendix: architectures}

We found our proposed deep learning architecture (Fig. \ref{fig:NN_scheme}) to perform best at predicting the gravity moments compared to other architectures tested. First, we tried a fully connected network architecture \citep{LeCun2015} with a varying depth and size (i.e., varying the number of hidden layers, with a varying number of neurons) predicting all five output parameters together. Secondly, we tried a similar fully connected architecture to regress each parameter separately, which was successful only for the mass prediction. Lastly, we adopted a sharing-based architecture \citep{Caruana2002}, again tested with different sizes and depths. Our chosen architecture may seem large compared for example with the network used by \cite{Agarwal2020} for a similar regression task, where they used three hidden layers with less than 100 neurons each. These authors compensated for the small network size with a very long training of \(4.4\times 10^6\) epochs, which is a few orders of magnitudes longer than our training process. Again, we note that no overfitting occurred during the training, supporting the validity of the architecture chosen.

\section{Quantitative DNN performance evaluation}\label{apendix: quantitative}

In addition to the performance evaluation shown in Fig. \ref{fig: accuarcies}, we present a quantitative comparison of the DNN mean prediction errors with other sources of uncertainty in Table \ref{tab:perfomance}. The \(1\sigma\) prediction errors on the validation dataset (\(\epsilon_{\sigma}\)) are comparable to the Juno measurement uncertainty for \(J_6\) and \(J_8\). All prediction errors are comparable to the uncertainty due to the wind, and they are significantly lower than the offsets needed to be applied on results produced by the   ToF expansion \citep{Zharkov1978}. The maximal absolute prediction errors (\(\epsilon_{\mathrm{max}}\)) are comparable to the offsets to ToF by \cite{Guillot2018}. In our CMS model, we set the outermost spheroid radius to be the measured \(R_{\mathrm{eq}}=71\,492\,\rm{km}\) \citep{Lindal1992} at one bar. This underlines the assumption that the higher atmosphere ($P < 1 \, \text{bar}$) can be neglected. \cite{Debras2018} evaluated the uncertainty stemming from this assumption, which is lower than \(\epsilon_{\sigma}\) for \(J_2\), but higher for \(J_4\) and \(J_6\). These authors also evaluated the discretization error on \(J_2\times10^6\), when using a polytropic EOS, to be of a similar magnitude to the value shown in Table \ref{tab:perfomance} for neglecting the higher atmosphere.

\begin{table*}
    \caption{NeuralCMS performance on the validation dataset compared to other sources of uncertainty.}
    \label{tab:perfomance}
    \centering
    \begin{tabular}{l c c c c c l}
    \hline\hline
     & \(J_{2}\times10^{6}\)\ & \(J_{4}\times10^{6}\) & \(J_{6}\times10^{6}\) & \(J_{8}\times10^{6}\)  & Mass \((10^{27}\rm{kg})\) & Reference\\
    \hline
        \(\epsilon_{\sigma}\) & 0.7890 & 0.0490 & 0.0045 & 0.0003 & 0.0001\\
        \(\epsilon_{\mathrm{max}}\) & 7.8550 & 0.5332 & 0.0964 & 0.0044 & 0.0020\\
        \hline
        Juno 3$\sigma$ & 0.0017 & 0.0024 & 0.0067 & 0.0207 & & \cite{Durante2020}\\
        \(GM_{\mathrm{Juno}}/G\) & & & & & 0.0005 & See Sect. \ref{sect2}\\
        Wind-related & 0.3543 & 0.0836 & 0.0768 & 0.0624 & & \cite{Miguel2022}\\
        Offsets to ToF magnitude & 1.7862 & 0.0605 & 0.0675 & 0.1679 & & \cite{Miguel2022}\\
        Offsets to ToF magnitude & 5.8554 & 0.4045 & 0.0375 & 0.1641 & & \cite{Guillot2018}\\
        CMS assumptions & \(\sim0.1\) & \(\sim0.06\) & \(\sim0.02\) & & & \cite{Debras2018}\\
    \hline
    \end{tabular}
    \tablefoot{CMS assumptions refer to neglecting the high atmosphere.}
\end{table*}

\section{Description of SHAP values}\label{apendix: shap}

The SHAP values are a game theory approach that assigns an impact value for each model input representing its contribution to a specific prediction \citep{Lundberg2017}. For deep learning models, SHAP is combined with the additive feature attribution DeepLIFT method \citep{shrikumar2017}, which practically linearizes the nonlinear components of the DNN, to provide explanations based on a locally accurate approximation: 
\[f(\Vec{x})\approx \bar{y}_{\mathrm{ref}}+\sum_{i=1}^{M}\phi_{x_{i},y},\]
where \(\Vec{x}\) is a specific input; \(f(\Vec{x})\) is the prediction model (the DNN in our case); \(\bar{y}_{\mathrm{ref}}\) is the mean of all output predictions \(y\) in a reference dataset, used as a baseline value; \(M\) is the number of input features; and \(\phi_{x_{i},y}\) is the SHAP value of the input \(x_i\) for a predicted output \(y\) \citep{Lundberg2017}. This means that for each prediction, the SHAP values for all inputs sum to the difference between the predicted value and the mean of all predictions in a reference dataset. For this work, we used Deep SHAP through the DeepExplainer from the SHAP Python library \citep{Lundberg2017}, which combines SHAP values computed for smaller components of the DNN into values for the whole network by backpropagating the computation through the network. We took all 2927 models accepted by their CMS results as the reference dataset and calculated SHAP values for 500 random models from these accepted models. More details on the specific model marked by black circles in Fig. \ref{fig:shap} can be found in Table \ref{tab:shap}.

\begin{table*}
    \caption{SHAP values for the prediction of \(J_6\) of an interior model with the highest SHAP value for \(Z_{\rm dilute}\) being shown with black circles in Fig. \ref{fig:shap}.}
    \label{tab:shap}
    \centering
    \begin{tabular}{l c c c c c c c c}
    \hline\hline
     & \(Y_{\rm{proto}}\) & \(T_{\rm{1bar}}\,(\rm{K})\) & \(Z_1\) & \(P_{12}\,(\rm{Mbar}\)) & \(m_{\rm dilute}\) & \(Z_{\rm dilute}\) & \(r_{\rm{core}}\,(r/R_{\rm{J}})\) & Sum \\
    \hline
        Parameter value & 0.28 & 183.87 & 0.02 & 2.57 & 0.60 & 0.07 & 0.10\\
        SHAP value (\(\delta J_{6}\times10^6\)) & $-0.19$ & $-0.35$ & 0.19 & 0.09 & $-1.28$ & 3.17 & $-1.76$ & $-0.13$\\
        \hline
    \end{tabular}
    \tablefoot{The prediction for this model is \(J_6\times10^{6}=34.22\), and the mean in the reference dataset is 34.35, showing that the sum of SHAP values approximates the difference between the prediction and the reference mean.}
\end{table*}

\end{appendix}

\end{document}